\documentstyle[preprint,aps]{revtex}
\begin{document}
\draft
\title{Quantum corrections to the conductivity of 
fermion--gauge field models: Application to half filled Landau level
and high-$T_c$ superconductors.}
\author{A.D. Mirlin$\sp{1,2}$  
 and  P. W\"{o}lfle$\sp{1}$}
\address{
$\sp{1}$ Institut f\"{u}r Theorie der Kondensierten Materie,
  Universit\"{a}t Karlsruhe, 76128 Karlsruhe, Germany} 
\address{
$^2$ Petersburg Nuclear Physics Institute, 188350 Gatchina, St.Petersburg, 
Russia.}
\date{\today}
\maketitle
\narrowtext
\tighten
\begin{abstract}
We calculate the Altshuler-Aronov type quantum correction to the
conductivity of $2d$ charge carriers in a random potential (or random
magnetic field) coupled to a transverse gauge field. The gauge fields
considered simulate the effect of the Coulomb interaction for the
fractional quantum Hall state at half filling and for the $t-J$ model
of high-$T_c$ superconducting compounds. 
We find an unusually large quantum correction
varying linearly or quadratically with the logarithm of temperature,
in different temperature regimes.  

\end{abstract}

\section{Introduction.}
\label{s1}

Recent experiments on the fractional quantum Hall state at
half-filling ($\nu=1/2$) \cite{rok} and on the high temperature 
superconductor
(HTSC) La$_{2-x}$Sr$_x$CuO$_2$ in a high magnetic field quenching the
superconducting state \cite{ando} have revealed an unexpected logarithmic
temperature dependence of the (longitudinal) resistivity $\rho_{xx}$. 
Similar behavior in HTSC compounds with low transition temperatures
has been reported earlier in low or vanishing magnetic field
\cite{preyer,fiory,jing}. There are several known sources of a
logarithmical rise of resistivity with decreasing temperature. The
first one, due to the Kondo effect \cite{kondo} of localized magnetic
spins, may be excluded here since the large magnetic fields employed
in the FQHE \cite{rok} and in the HTSC \cite{ando} experiments would
quench the Kondo effect up to a rather high energy scale. This
eliminates the possibility of a $\ln T$ dependence at the observed low
temperatures. Also, there is no indication of the existence of local
moments in these systems. The second possible source is weak
localisation (WL) of charge carriers in a $2d$ disordered
potential. However, the large magnetic fields used in the experiments
\cite{rok,ando} destroy the quantum coherent backscattering effect
responsible for WL. 
WL may contribute to the $\ln T$ behavior
observed in HTSC in zero magnetic field \cite{preyer,fiory,jing},
though. However, the magnitude of the $\ln
T$ term found in these experiments is too large to be explained solely
by the WL. The third effect giving rise to $\Delta\rho\sim\ln(1/T)$ at
low temperature $T$ derives from the quantum interference of impurity
scattering and interaction between the charge carriers. Roughly
speaking, the diffusive rather than ballistic motion of charge
carriers in a random potential increases the time two interacting
particles spend near each other and enhances therefore the interaction
strength. In $2d$ the correction even diverges, eventually leading to
the localization of the particles at zero temperature. The effect was
discovered by Altshuler and Aronov (AA) \cite{AA}, who obtained the
quantum correction to the conductivity
$\Delta\sigma=(e^2/h)\lambda_{AA}\ln(T\tau)$. Here $e^2/h$ is the
quantum unit of conductance, $\tau$ is the lifetime due to elastic
scattering, and $\lambda_{AA}=\pi^{-1}(1-{3\over 2}F)$ is a numerical
coefficient. The quantity $F$ is due to Hartree processes and is equal
to the exact static scattering amplitude of a particle and a hole with
total spin $j=1$. It is usually positive, leading to partial
cancellation of the exchange contribution represented by the first
term (unity) in brackets,and tends to zero in the limit
of weak screening. As we will discuss in detail below, the magnitude
of the logarithmic term observed experimentally in HTSC materials and
at $\nu=1/2$ is typically a factor of $3-5$ larger than the AA
result. As we will show, enhancement factors of this order may
be obtained if the interaction by a fictitious gauge field is considered
in addition to the usual Coulomb interaction. 

The problem of particles interacting with a transverse gauge field
was studied by Reizer \cite{reizer} in the context of magnetic interaction
of electrons in metals. More recently, its 2d version
 has attracted  considerable
interest \cite{gauge} in connection with  the gauge theory of high-$T_c$
superconductors \cite{htsc}
and the Chern-Simons theory of the half-filled
Landau level \cite{lofra,hlr}. 
Both of these theories  have definite successes in explaining 
the experimental situation and seem to capture at least the essential
physics of the phenomena. This has motivated us to study in detail the
low-temperature correction to conductivity in the framework of the above
gauge theories. In the context of half-filled Landau level such a
calculation was very recently done by Khveshchenko \cite{khve}. We find however
that his results are incomplete (some of the temperature regimes are
missing, some relevant parameters are not introduced) and contain some
errors in the coefficients, which are crucial for 
comparison to the experimental data.

\section{Quantum correction to conductivity of fermions coupled to a
transverse gauge field}
\label{s1a}

We start from considering a simplified model of spinless
fermions subject to a static random potential/magnetic field and coupled to a
transverse gauge field, in the absense of any other
interactions. Complications which arise in the cases of composite
fermions at $\nu=1/2$ and the gauge theory of high-$T_c$ compounds will be
discussed in the next two sections.
The transverse part of the gauge field
propagator is determined by the polarization tensor of fermions and
has the form\cite{reizer,gauge}
\begin{equation}
D_{\mu\nu}(\bbox{q},\omega)={\delta_{\mu\nu}-\hat{q}_\mu\hat{q}_\nu \over
i\omega\sigma(q,\omega) + \chi q^2}\ ,
\label{e1}
\end{equation}
where $\bbox{\hat{q}}=\bbox{q}/|\bbox{q}|$, and  $\chi=1/(24\pi m)$
is the diamagnetic susceptibility, with $m$ being the fermion mass.
The conductivity $\sigma(q,\omega)$ may be
approximated by its static, long wavelength limit,
$\sigma_0=k_Fl/4\pi$, with $k_F=(4\pi n)^{1/2}$ the Fermi wave number
and $l$ the transport mean free path.

The quantum
correction to the conductivity is calculated following Altshuler and
Aronov in lowest order of perturbation theory in the gauge field
interaction for charge carriers diffusing in a random potential and/or
random magnetic field. Vertex corrections due to impurity scattering
are responsible for the singular enhancement of the interaction and
are included consistently in lowest order in the disorder parameter 
$1/(E_F\tau_{tr})$, where $E_F$ is the Fermi energy and $\tau_{tr}$ is
the momentum (transport) relaxation time due to impurity scattering.
    
The two leading diagrams for the conductivity are shown in
Fig.1. Three more diagrams included by AA are not relevant in the case
of a transverse gauge field considered here, since they are not affected
by diffusion poles. The solid lines in Fig.1 denote Green's functions 
$$
G(\bbox{k},\epsilon_n)=[i\epsilon_n-\xi_k+(i/2\tau_k)sign(\epsilon_n)]^{-1}
\ , 
$$
where $\epsilon_n=(2n+1)\pi T$ is the Matsubara energy, $\xi_k=v_F(k-k_F)$
is the quasiparticle energy for momentum $\bbox{k}$ and $\tau_k$ is
the single particle relaxation time. As we shall show, the single
particle time drops out of the final result, so we will not be
concerned about the difficulties encountered with the definition of
$\tau_k$ in the case of random magnetic fields \cite{amw,aamw}. The
wavy line in Fig.1 represents the gauge field propagator,
$D_{\mu\nu}(\bbox{q},\omega)$,
and the shaded blocks denote ``diffusons'' (particle-hole
propagators), given by \cite{bwr}
\begin{equation}
\Gamma_{kk'}(\bbox{q},\omega_m)=(2\pi
N_0)^{-1}{\gamma_k\gamma_{k'}^*\over\omega_m+Dq^2}
+\Gamma_{kk'}^{reg}\ ,
\label{e3}
\end{equation}
assuming the energies of a particle and a hole have opposite signs. 
Here $N_0$ is the density of states at the Fermi level, $D$ is the
diffusion coefficient, and
$$
\gamma_k={1\over\tau_k}-i\langle\bbox{q}\bbox{v}_{k_1}\tau_{k_1}
\Gamma_{kk_1}^{reg}\rangle_{k_1}\ ,
$$ 
where $\langle\ldots\rangle_k$ denotes averaging over the Fermi
surface.
At small frequency and momentum, $\Gamma_{kk'}$ is dominated by the
singular contribution, which is given in the case of spatial isotropy,
$\tau_k=\tau$, by
\begin{equation}
\Gamma_{kk'}(q,\omega_m)\simeq {1\over 2\pi N_0\tau^2(\omega_m+Dq^2)}
\label{e4}
\end{equation}
The regular part  $\Gamma^{reg}$ is however also important, since it
determines the vertex renormalization, denoted by shaded triangles on
Fig.1. $\Gamma^{reg}$ may be expressed in terms of the
eigenvalues $\lambda_l$ and eigenfunctions $\Phi_k^l$ of the bare
scattering cross section $w_{kk'}$ as
$$
\Gamma_{kk'}^{reg}=\sum_{l\ne 0}{\lambda_l\over
1-\lambda_l}\Phi_k^l\Phi_{k'}^{l*}\ ,
$$
excluding the eigenvalue $\lambda_0=1$ associated with  particle number 
conservation. The diffusion coefficient is defined in this framework
as
\begin{equation}
D=\langle\tau_k v_{k\mu}^2\rangle_k+\langle
v_{k\mu}\tau_k\Gamma^{reg}_{kk'}\tau_{k'} v_{k'\mu}\rangle_{k,k'}
\label{e5}
\end{equation}
In the case of spatial isotropy $\Phi_k^l=(2\pi)^{-1/2}e^{il\phi_k}$
($\phi_k$ is the polar angle of the momentum $k$), and eq.(\ref{e5})
reduces to
\begin{equation}
D={1\over 2} v_F^2\tau_{tr}\ ;\qquad \tau_{tr}={\tau\over 1-\lambda_1}
\label{e6}
\end{equation}
The velocity vertex $v_{k\mu}$ renormalizes to
\begin{equation}
J_{k\mu}=v_{k\mu}+\sum_{l\ne 0}{\lambda_l\over 1-\lambda_l}\langle
v_{k'\mu}\tau_{k'} \Phi_{k'}^l\rangle_{k'}\Phi_{k}^l=v_{k\mu}{1\over
1-\lambda_1} =v_{k\mu} {\tau_{tr}\over\tau}
\label{e7}
\end{equation}
This applies  when the energy sign is opposite for two fermion
lines attached to the vertex; otherwise, the vertex retains its bare
form $v_{k\mu}$. 

Now we calculate the conductivity correction represented by the two
diagrams of Fig.1 (plus the same diagrams with reversed arrows) in the
Matsubara technique. Each of the two fermion loops  produces the
factor $\pm 2\pi iN_0\tau^2 {v_F^2\over 2}$. The diffuson is approximated by
eq.(\ref{e4}) provided the two fermion lines it connects have
 opposite energy signs. This requirement implies that exactly two of
four velocity vertices get renormalized according to eq.(\ref{e7}). 
Combining all this, we find that the total relaxation rate $\tau$
drops out, and the conductivity correction at a Matsubara frequency
$\Omega_k>0$ takes the following form:  
\begin{equation}
\Delta\sigma(i\Omega_k,T)=-e^2{4\over \Omega_k} N_0
D^2T\left\{\sum_{\omega_l=0}^{\Omega_k}\omega_l
F(i\Omega_k,i\omega_l)+\sum_{\omega_l>\Omega_k}\Omega_k
F(i\Omega_k,i\omega_l) \right\}\ ,
\label{e8}
\end{equation}
where
\begin{equation}
F(i\Omega_k,i\omega_l)=\int {d^2q\over (2\pi)^2}{1\over
(Dq^2+\Omega_k+\omega_l) (\omega_l\sigma(q)+\chi q^2)}
\label{e9}
\end{equation}
Carrying out the analytical continuation to  real values of the
frequency $\Omega$, we get
\begin{equation}
\Delta\sigma(\Omega,T)= e^2{N_0D^2\over
i\pi\Omega}\int_{-\infty}^{\infty} d\omega\,
\omega\coth\left({\omega\over
2T}\right)[F(\Omega,\omega+\Omega)-F(\Omega,\omega)]\ ,
\label{e10}
\end{equation}
with
\begin{equation}
F(\Omega,\omega)=\int {d^2q\over (2\pi)^2}{1\over
(Dq^2-i\Omega-i\omega) (-i\omega\sigma(q)+\chi q^2)}
\label{e11}
\end{equation}
Now we are going to evaluate eqs.(\ref{e10}), (\ref{e11}) at zero
external frequency $\Omega$, when eq.(\ref{e11}) can be rewritten as
\begin{eqnarray}
\Delta\sigma(T)&=&-{e^2N_0D^2\over i\pi}\int_{-\infty}^{\infty}d\omega
F(0,\omega) {\partial\over\partial\omega}\left[\omega\coth{\omega\over
2T}\right] \nonumber\\
&\simeq&-\mbox{Im}{2e^2N_0D^2\over
\pi}\int_T^{1/\tau_{tr}}F(0,\omega)d\omega 
\label{e12}
\end{eqnarray}
The integrand in eq.(\ref{e11}) has at $\Omega=0$ 
three characteristic momentum scales, where its behavior changes:
\begin{itemize}
\item[i)] $q_D=\sqrt{\omega/D}=l^{-1}\sqrt{2\omega\tau_{tr}}$, 
determined by the first factor in the
denominator;
\item[ii)] $q_\chi=
\sqrt{\omega\sigma_0/\chi}=k_F\sqrt{6\omega\tau_{tr}}$, 
determined by
the second factor in the denominator;
\item[iii)] $l^{-1}$, which limits the applicability of the diffusion
approximation.
\end{itemize}
Comparing these three scales, we see that $q_D$ is the smallest one,
whereas the relation between the other two depends on the value of
frequency $\omega$. We distinguish therefore the following
 two frequency domains:
\begin{itemize}
\item[i)] $\omega\ll T_0\equiv\displaystyle{ 
 {1\over 6(k_Fl)^2\tau_{tr}} }$.\\
In this region $q_\chi\ll l^{-1}$, and the integral (\ref{e11}) yields
\begin{eqnarray}
F(0,\omega)&\simeq&{1\over 2\pi}\int_0^\infty{qdq\over
(Dq^2-i\omega)(\chi q^2-i\omega\sigma_0)}\simeq{i\over 4\pi
D\sigma_0\omega}\ln{D\sigma_0\over\chi} \nonumber\\
&=&{i\over 2\pi N_0D^2\omega}\ln(\sqrt{3}k_Fl)
\label{e23a}
\end{eqnarray}
\item[ii)] $\omega\gg T_0$.\\
In this case,  $l^{-1}\ll q_\chi$, yielding
\begin{equation}
F(0,\omega)\simeq{i\over 2\pi N_0D^2\omega}\ln{1\over lq_D}=
{i\over 4\pi N_0D^2\omega}\ln{1\over\omega\tau_{tr}}
\label{e24}
\end{equation}
\end{itemize}
Substituting eqs.(\ref{e23a}), (\ref{e24}) in eq.(\ref{e12}), we get  
the following result for the correction to the conductivity
\begin{equation}
\Delta\sigma(T)={e^2\over 2\pi h}\times\left\{
\begin{array}{ll}
\displaystyle{ 4\ln(\sqrt{3}k_Fl) \ln (T\tau_{tr}) }\ ,& \qquad T\ll T_0 \\
\displaystyle{ -\ln^2 (T\tau_{tr}) }\ ,&\qquad T_0\ll
T\ll 1/\tau_{tr}
\end{array}
\right.
\label{e25a}
\end{equation}
We see that at low temperatures the correction has the familiar $\ln
T$ form, but with the coefficient enhanced by a large factor
$2\ln(\sqrt{3}k_Fl)$ as compared to the usual Altshuler-Aronov result
\cite{AA}.

\section{Interaction correction to the conductivity of composite
fermions at half filling of the Landau level.}
\label{s2}

The longitudinal resistivity $\rho_{xx}$ 
of a Hall bar in a strong magnetic field
shows oscillations with minima at filling factors $\nu=p/(2p+1)$ of
the lowest Landau level, where $p$ is a positive or negative
integer. For not too large values of $p$, $|p|=1,2,3$, the
resistivity is exactly zero at the minima and the Hall resistivity is
quantized: $\rho_{xy}={2p+1\over p}{h\over e^2}$. In the opposite
limit $|p|\to\infty$, or $\nu\to 1/2$, $\rho_{xx}$ tends to a finite
limiting value, which is approached in a form of decaying
oscillations, reminiscent of Shubnikov--de Haas oscillations in weak
magnetic fields. This and other observations has been interpreted as
indicating a kind of Fermi liquid state near $\nu=1/2$. As a possible
theoretical framework for this phenomenon, Jain \cite{jain} has
proposed the composite fermion picture, in which a physical electron
is replaced by a fermion threaded by a flux-tube carrying two flux
quanta oriented oppositely to the external magnetic field $B$. The mean
field generated by the flux tubes reduces the applied magnetic field
to the effective field $B_{eff}=B-4\pi cn/e$ ($n$ is the electron
density), which vanishes at half-filling. The quantized values of
$\nu$ correspond to integer filling factors for the Landau levels in
the effective fields, providing a natural explanation for the
dominance of the observed plateau values.

A field-theoretical formalism implementing this idea with the help 
of a Chern-Simons (CS) gauge field has been worked out by Lopez and
Fradkin \cite{lofra}. Following a similar approach, Halperin, Lee and
Read \cite{hlr} have developed a theory for the half-filled Landau
level, in which the fluctuations of the gauge field about the mean
field are treated in the Random Phase Approximation (RPA). At this
level of approximation, the transverse part of the gauge field
propagator has the form (\ref{e1}).
 Here the renormalized diamagnetic
susceptibility is given by \cite{hlr,maw}
\begin{equation}
\chi={1\over 2\pi m^*}\left[{1\over
12}+\left(2\pi\sigma_{xy}^f+{1\over 2} \right)^2\right]+{v(q)\over
(4\pi)^2}\ ,
\label{e2}
\end{equation}
where $\sigma_{xy}^f$ is the Hall conductivity of composite fermions,
which vanishes at $\nu=1/2$, and $v(q)$ is the Coulomb interaction
$v(q)=2\pi e^2/\epsilon(q+\kappa)$, with $\epsilon$ being the
dielectric constant. The screening parameter $\kappa$ is
governed by the polarization of external charges in gates, leads,
doping layers, etc.  

\subsection{Unscreened Coulomb interaction.}
\label{s2.1}

In the case of unscreened Coulomb interaction, $v(q)=2\pi e^2/\epsilon
q$, the effective susceptibility $\chi$ is determined by the last term
in eq.(\ref{e2}):
\begin{equation}
\chi\simeq{e^2\over 8\pi\epsilon q}
\label{e13}
\end{equation}
The momentum scale $q_\chi$ is now given by
$$
q_\chi=\displaystyle{
{2\epsilon k_F l\over e^2}\omega }
$$
Comparing it to $q_D$ and $l^{-1}$, we find that one should distinguish
now between the three following regions of frequency:
\begin{itemize}
\item[i)] $\omega\ll T_1\equiv\displaystyle{
{C_*^2\over 2(k_Fl)^2\tau_{tr}}}$,\\
where $C_*=e^2m^*/(\epsilon k_F)$ is a numerical constant, which
is of order of 10 according to the experimental data \cite{cf}.  In this
low-frequency region we have $q_\chi\ll q_D\ll l^{-1}$, and the
integral (\ref{e11}) can be estimated as
\begin{equation}
F(0,\omega)\simeq{1\over 2\pi}{8\pi\epsilon\over
e^2}\int_0^\infty{dq\over Dq^2-i\omega}={2\pi\over C_*
v_F^2}\left({2i\over\omega\tau_{tr}}\right)^{1/2}
\label{e14}
\end{equation}
\item[ii)] $T_1\ll \omega\ll T_2\equiv
\displaystyle{ {C_*\over 2k_Fl\tau_{tr}} }$. \\
In this domain $q_D\ll q_\chi\ll l^{-1}$, and we find
\begin{eqnarray}
F(0,\omega)&\simeq&{1\over 2\pi}\int {qdq\over
\left(-i\omega N_0D+{e^2\over 8\pi\epsilon}q\right)(Dq^2-i\omega)}
\nonumber\\
&\simeq&{i\over 4\pi N_0D^2\omega}\ln\left[{2\over
C_*^2}(k_Fl)^2\omega\tau_{tr}\right] 
\label{e15}
\end{eqnarray}
\item[iii)] $\omega\gg T_2$.\\
Now we have $q_D\ll l^{-1}\ll q_\chi$, and $F(0,\omega)$ is given by
eq.(\ref{e24}). 
\end{itemize}

Evaluating finally the frequency integral in eq.(\ref{e12}), we find
the following correction to the conductivity
\begin{equation}
\Delta\sigma(T)={e^2\over 2\pi h}\times\left\{
\begin{array}{ll}
\displaystyle{ 2\pi{ k_Fl\over C_*}\sqrt{T\tau_{tr}} }\ ,& \qquad T\ll T_1 \\
\displaystyle{ -\ln^2\left[{2(k_Fl)^2\over
C_*^2}T\tau_{tr}\right]} \
,&\qquad T_1\ll T\ll T_2 \\
\displaystyle{ -\ln^2 (T\tau_{tr}) }\ ,&\qquad T_2\ll
T\ll 1/\tau_{tr}
\end{array}
\right.
\label{e17}
\end{equation}

\subsection{Screened Coulomb interaction}
\label{s2.2}

Now we will suppose the Coulomb interaction to be screened with a
short enough screening length $\kappa^{-1}\lesssim l$, so that its Fourier
transform can be replaced by a constant
\begin{equation}
v(q)\simeq{2\pi e^2\over\epsilon\kappa}={2\pi\over
m^*}C_*{k_F\over\kappa}
\label{e18}
\end{equation}
The susceptibility $\chi$ is then  according to eq.(\ref{e2}) given by
\begin{equation}
\chi={1\over 2\pi m^*}\left({1\over 12}+{1\over 4}+{1\over
4}C_*{k_F\over\kappa} \right)
\label{e19}
\end{equation}
Taking into account that $C_*\sim 10\gg 1$ and $k_F/\kappa\gtrsim 1$,
we can neglect all but the last term in eq.(\ref{e19}), so that
\begin{equation}
\chi\simeq {C_*\over 8\pi m^*}{k_F\over\kappa}
\label{e20}
\end{equation}
The scale $q_\chi$ is now given by
\begin{equation}
q_\chi=\left({\omega\sigma_0\over\chi}\right)^{1/2}= 
\left(2{\kappa\over k_F}{k_Fl\over C_*}m^*\omega\right)^{1/2}
\label{e21}
\end{equation}
We will assume that
\begin{equation}
\left({q_\chi\over q_D}\right)^2={\kappa\over k_F C_*}(k_Fl)^2>1\ ,
\label{e22}
\end{equation}
so that $q_D$ is the smallest of the three scales $q_D$, $q_\chi$ and
$l^{-1}$. As in section \ref{s1a}, we find then two frequency domains:
\begin{itemize}
\item[i)] $\omega\ll T_0^*\equiv\displaystyle{ {C_*\over 2}
{k_F\over\kappa} {1\over (k_Fl)^2\tau_{tr}} }$.\\
In this case $q_D\ll q_\chi\ll l^{-1}$, and we get
\begin{eqnarray}
F(0,\omega)&\simeq&{1\over 2\pi}\int_0^\infty{qdq\over
(Dq^2-i\omega)(\chi q^2-i\sigma_0)}\simeq{i\over 4\pi
N_0D^2\omega}\ln{N_0D^2\over\chi} \nonumber\\
&=&{i\over 4\pi N_0D^2\omega}\ln\left[{(k_Fl)^2\over C_*}{\kappa\over
k_F}\right] 
\label{e23}
\end{eqnarray}
\item[ii)] $\omega\gg T_0^*$.\\
In this region $q_D\ll l^{-1}\ll q_\chi$, and $F(0,\omega)$ is given
by eq.(\ref{e24}). 
\end{itemize}
We find therefore the following correction to the conductivity
\begin{equation}
\Delta\sigma(T)={e^2\over 2\pi h}\times\left\{
\begin{array}{ll}
\displaystyle{ 4\ln\left[\left({\kappa\over
C_*k_F}\right)^{1/2}k_Fl\right] \ln (T\tau_{tr}) }\ ,& \qquad T\ll T_0^* \\
\displaystyle{ -\ln^2 (T\tau_{tr}) }\ ,&\qquad T_0^*\ll
T\ll 1/\tau_{tr}
\end{array}
\right.
\label{e25}
\end{equation}

Let us estimate the magnitude of the low-temperature logarithmic
correction. We will take for this purpose the typical values of the
parameters $k_Fl=2(h/e^2)\sigma_{xx}^f\simeq 50$ \cite{rok} and
$C_*\simeq 10$ \cite{cf}. Further, we assume the screening length
$\kappa^{-1}$ to be of order of the spacer distance $d_s=120 nm$, so
that $k_F/\kappa\sim 10$. Then the coefficient $\lambda$ of the
low-temperature correction
\begin{equation}
\Delta\sigma(T)={e^2\over h}\lambda\ln(T\tau_{tr})
\label{e26}
\end{equation}
can be estimated as $\lambda\simeq{2\over\pi}\ln 5\simeq 1.0$. This
value is in good agreement with the experimental results,
$0.4\le\lambda\le 1.6$ reported in Ref.\cite{rok}. We should note,
however, that the transport mean free time $\tau_{tr}$ and the 
crossover temperature $T_0^*$  are estimated 
with these values of parameters as $1/\tau_{tr}\sim 0.5 K$ and 
$T_0^*\sim 10-20 mK$, so that the experimental data of Ref.\cite{rok}
belong mainly to the high-temperature region (second line of
eq.(\ref{e25})). Though the slope $d\sigma/d\ln T$ of the corresponding
log-squared behavior is the same at $T=T_0^*$ as for the simple logarithmic
behavior at $T<T_0^*$, it is expected to decrease according to
eq.(\ref{e25}) with $T$ approaching $1/\tau_{tr}$. This would lead to a
curvature of the graph  $\sigma(\ln T)$; a feature, which seems
not be observed in the experiment \cite{rok}. The reason of this
discrepancy is not clear to us at the present stage. 

\section{Interaction correction to conductivity in HTSC materials.}

The properties of the high-$T_c$ superconducting compounds are
governed by the physics of the quasi-two-dimensional system of
CuO$_2$-planes. A good candidate model is the $t-J$--Hamiltonian of
two-dimensional electrons interacting via a nearest neighbor spin
exchange interaction $J$ and hopping on a square lattice with nearest 
neighbor amplitude $t$. Only the regime of small hole doping is of
interest. The hopping is constrained to singly occupied lattice sites,
on account of the large on site Coulomb repulsion on the Cu
sites. It is difficult to implement the constraint in analytic
approximations. A widely used method is the slave boson approach, in
which bosons ($b$) are introduced to describe the empty lattice sites,
in addition to two fermion ($f$) species with spin up and down. The
constraint takes the form $b_i^\dagger b_i+\sum_\sigma
f_{i\sigma}^\dagger f_{i\sigma}=1$ for each lattice site $i$. A mean
field approximation using the coherent hopping amplitudes 
$\langle b_i^\dagger b_i \rangle$, 
$\langle f_{i\sigma}^\dagger f_{i\sigma}\rangle$,
the boson condensation amplitude $\langle b_i \rangle$, and the
singlet pair amplitude $\langle f_{i\uparrow}f_{i\downarrow}\rangle$ as
collective fields yields a phase diagram, which remarkably resembles
experimental observations. We are here interested in the normal state
properties at optimal doping, which is the phase where 
$\langle b_i^\dagger b_i \rangle$ and $\langle f_{i\sigma}^\dagger
f_{i\sigma}\rangle$ are finite, while $\langle b_i \rangle$ and
$\langle f_{i\uparrow}f_{i\downarrow}\rangle$ are zero (strange metal
phase). In this phase the low energy
fluctuations about the mean field may be 
described in terms of a $U(1)$ gauge field, introduced to restore the
local constraint violated in mean field theory. The properties of the
effective fermion--boson--gauge field theory have been discussed by
Ioffe and Larkin and by Lee and Nagaosa among others \cite{htsc}. 
The gauge field propagator has the form (\ref{e1}), with the
conductivity $\sigma$ and the susceptibility $\chi$ given now by
the sum of fermion and boson contributions:
$\sigma=\sigma_b+\sigma_f$, $\chi=\chi_b+\chi_f$. Since the
concentration of bosons is equal to that of the holes and is therefore
small for low doping, it is believed that $\sigma_b \ll \sigma_f$, so
that the gauge field propagator (\ref{e1}) is determined essentially
by the fermion polarization. For the fermion conductivity $\sigma_f$
we could use therefore the above analysis of Sec.\ref{s1a} with the
final result (\ref{e25a}). However, according to the Ioffe-Larkin
composition rule, the physical conductivity (response to an external
electromagnetic field) is given not by  
$\sigma=\sigma_b+\sigma_f$, but rather by
$\sigma_{ph}=(\sigma_b^{-1}+\sigma_{f}^{-1})^{-1}$. 
Therefore, for $\sigma_b \ll \sigma_f$ we have
$\sigma_{ph}\approx\sigma_b$, so that physical conductivity
is dominated by bosons. One can argue that the repulsive interaction
prevents them from Bose-condensation (except possibly at very low
temperatures), and the bosons form a kind of Fermi surface.
Then our above  consideration in Sec.\ref{s1a}
can still be applied. However, since the diffusion coefficient $D$ in
eqs.(\ref{e10}), (\ref{e11})  refers now to bosons, whereas the
conductivity $\sigma$ entering the gauge field propagator corresponds
to fermions, we  get an extra small overall factor
$\sigma_b/\sigma_f$ in the  final result, eq.(\ref{e25a}). 
In addition, the factor $k_Fl$ in the argument of the logarithm is
replaced by $(k_Blk_Fl_f)^{1/2}$, where $k_B$ is a characteristic
momentum of the bosons, and  $l_f$ and $l$ are fermion
and boson mean free paths, respectively.

Let us compare these results with  experimental findings. In
ref.\cite{preyer} data for the value of the coefficient $\lambda$ of the
logarithmic correction eq.(\ref{e26}) are presented for LaSrCuO
samples with values of $k_Fl$ ranging from $1$ to $6$. The
corresponding $\lambda$'s lie in the interval 0.8--1.8 and show a
tendency of slight increase with the value of $k_Fl$. These results are
in a surprisingly good agreement with our eq.(\ref{e25a}), but without
the extra factor $\sigma_b/\sigma_f$. Similar conclusion can be drawn
from the analysis of other experimental data \cite{ando,fiory,jing}.

We are thus forced to conclude that in the present state of the gauge
theory of high-$T_c$ materials, the magnitude of the correction to the
physical conductivity due to the exchange of the gauge field is much
smaller that its experimentally measured value, because of the small
factor $\sigma_b/\sigma_f$. However, taking into account the rather
good agreement between  the experiment and the theoretical result
(\ref{e25a}) without this factor, it is tempting to speculate about
possible modification of the gauge theory. The factor
$\sigma_b/\sigma_f$ would not appear, if $\sigma_f(q)$ would fall
below $\sigma_b(q)\approx\sigma_b(0)$ for most of the $q$-regime of
interest ($l_f^{-1}\ll q\ll l^{-1}$). This would require the
characteristic momentum $k_B$ in $\sigma_b\sim k_Bl$ to be much larger
than the Fermi momentum $k_F$. Alternatively, the effective gauge
theory might include only one type of particles.

\section{Conclusion.}

We have calculated the quantum correction to the conductivity of
charge carriers in a two-dimensional disordered system subject to a
transverse gauge field interaction. We find the $\ln T$ behavior at
low temperature, but with a coefficient strongly enhanced as compared
to the usual Altshuler-Aronov result. At higher temperatures, a
crossover to $\ln^2T$ behavior is found. The results are in good
quantitative agreement with the recently
observed $\ln T$ dependence of the resistivity at half filling of the
Landau level. We have also discussed the possible relevance of our
results to the low temperature resistivity of high-$T_c$
superconducting compounds in the framework of a gauge field model.

\section{Acknowledgements.}

We acknowledge useful discussions with D.V.Khveshchenko, P.A.Lee and
H.Fukuyama. This work was supported by Sonderforschungsbereich 195 der
Deutschen Forschungsgemeinschaft.

\begin{figure}
\caption{Diagrams for the correction to conductivity due to a
transverse gauge field interaction.}
\end{figure}

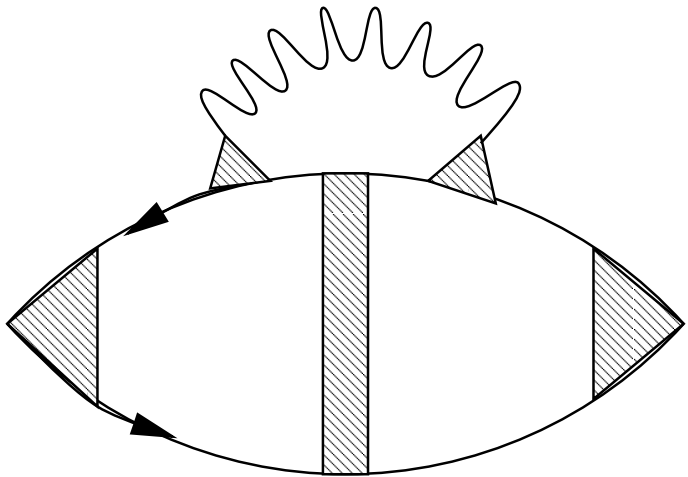
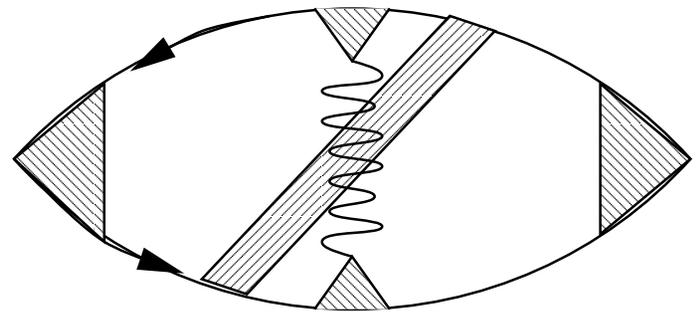
\end{document}